\pdfoutput=1
\documentclass[prd,preprintnumbers,nofootinbib,superscriptaddress,onecolumn,preprint]{revtex4-1}

\pdfoutput=1
\usepackage{graphicx}
\usepackage{amsmath}
\usepackage{amssymb}
\usepackage{slashed}
\usepackage{mathtools}
\usepackage[utf8]{inputenc}
\usepackage[colorlinks=true,linkcolor=blue,bookmarksopen,bookmarksnumbered]{hyperref}
\usepackage{color}
\usepackage[usenames,dvipsnames]{xcolor}
\usepackage[normalem]{ulem}
\usepackage{soul}
\usepackage{units}
\usepackage{rotating}
\usepackage{hhline,multirow,tabularx}
\usepackage{epstopdf}

\newcommand{\T}[1]{\boldsymbol{#1}_{\text{T}}}

\newcommand\3[1]{\boldsymbol{#1}}

\newcommand{\Tscsq}[2]{#1^2_{#2\text{T}}}
\newcommand{\no}{\nonumber \\}
\newcommand{\parz}[1]{\ensuremath{\left(#1\right)}}
\newcommand{\order}[1]{\ensuremath{O\parz{#1}}}
\newcommand{\mhad}{\ensuremath{M}}
\newcommand{\pmass}{\ensuremath{M}}

\newcommand{\xbj}{\ensuremath{x_{\rm Bj}}}
\newcommand{\xn}{\ensuremath{x_{\rm N}}}

\newcommand{\diff}[1]{\mathrm{d}#1}

\newcommand{\contractortot}[1]{\ensuremath{{\rm P}}_{#1}}

\newcommand{\xsub}{\ensuremath{X}}
\newcommand{\xbjh}{\ensuremath{\hat{x}_{\rm Bj}}}
\newcommand{\xnh}{\ensuremath{\hat{x}_{\rm N}}}
\newcommand{\mtsq}{\ensuremath{\Tscsq{m}{}}}
\newcommand{\xfunc}{\ensuremath{\mathcal{F}}}
\newcommand{\finmass}{\ensuremath{v}}
\newcommand{\finmassex}{\ensuremath{\tilde{\finmass}}}
\newcommand{\findelt}{\ensuremath{\delta \finmass}}

\newcommand{\eref}[1]{Eq.~(\ref{e.#1})}
\newcommand{\erefs}[2]{Eqs.~(\ref{e.#1})--(\ref{e.#2})}
\newcommand{\fref}[1]{Fig.~\ref{f.#1}}

\newcommand{\aref}[1]{Appendix~\ref{a.#1}}
\newcommand{\sref}[1]{Sec.~\ref{s.#1}}

\begin{document}

\title{What does kinematical target mass sensitivity in DIS \\
	reveal about hadron structure?}

\preprint{JLAB-THY-19-2880}
\author{E.~Moffat}
\email{emoff003@odu.edu}
\affiliation{Department of Physics, Old Dominion University, Norfolk, VA 23529, USA}
\author{T.~C.~Rogers}
\email{tedconantrogers@gmail.com}
\affiliation{Department of Physics, Old Dominion University, Norfolk, VA 23529, USA}
\affiliation{Jefferson Lab, Newport News, VA 23606, USA}
\author{W.~Melnitchouk}
\affiliation{Jefferson Lab, Newport News, VA 23606, USA}
\author{N.~Sato}
\affiliation{Department of Physics, Old Dominion University, Norfolk, VA 23529, USA}
\affiliation{Jefferson Lab, Newport News, VA 23606, USA}
\author{F.~Steffens}
\address{Instit\"ut f\"ur Strahlen- und Kernphysik, Universit\"at Bonn, Nussallee 14-16, 53115 Bonn, Germany}

\date{\today}

\begin{abstract}

We study the role of purely external kinematical approximations
in inclusive deep-inelastic lepton--hadron scattering within QCD
factorization, and consider factorization with an exact treatment
of the target hadron mass.
We discuss how an observed phenomenological improvement obtained
by accounting for target mass kinematics could be interpreted in
terms of general properties of target structure, and argue that
such an improvement implies a hierarchy of nonperturbative scales
within the hadron.

\end{abstract}

\maketitle

\section{Introduction}
\label{s.intro}

Understanding the nature of confined systems of strongly interacting
quarks and gluons (or partons), such as hadrons and nuclei, remains
one of the single most challenging problems in nuclear and particle
physics.
An essential tool in this quest has been the factorization of the
short- and long-distance parts of scattering amplitudes, which has
allowed the systematic study of hard scattering processes in terms
of universal sets of parton distribution functions (PDFs)
\cite{Collins:1988gx}.  While this has proved an enormously
successful paradigm when applied to reactions at high energies,
where typical momentum transfers $Q$ are much greater than any
hadronic mass scales, $Q \gg O(1~{\rm GeV})$, delineating the
extent to which factorization techniques may be applicable at
lower energies has been a rather more formidable task.

The transition region at intermediate momentum transfers,
$Q \sim 1$ -- 2~GeV, where descriptions of phenomena in terms
of parton degrees of freedom give way to nonperturbative
dynamics, is still poorly understood.
Here small-coupling quantum chromodynamics (QCD) techniques
are often applicable, while at the same time hadronic mass
effects are not always negligible.
Phenomena specific to the this regime, such as quark-hadron
duality~\cite{Bloom:1971ye, DeRujula:1976baf, Poggio:1975af,
Ji:1994br, Melnitchouk:2005zr} and precocious
scaling~\cite{Duke:1979na, Devoto:1983sh} have attracted
much interest, and it has been the focus of dedicated
experimental efforts at Jefferson
Lab~\cite{Christy:2011cv, Chen:2011zzp, Gilman:2011zz}.

Extending standard perturbative QCD (and even general partonic pictures)
to the \mbox{low-$Q$} region also presents theoretical challenges,
particularly since certain mass effects that are normally treated
as negligible in QCD processes with large $Q$ may become important
there.  Consider the basic statement of factorization for the inclusive
lepton--nucleon (or any other hadron or nucleus) DIS cross section
(see, {\it e.g.}, Ref.~\cite{Collins:1988gx}),
\begin{equation}
\frac{\diff{\sigma}}{\diff{\xbj}{} \diff{Q^2}{}}
= \int\!\diff{\xi}\,
  \frac{\diff{\hat{\sigma}}} {\diff{\xbjh}{} \diff{Q^2}{}}\, f(\xi;Q)
+ {\rm p.s.}\, ,
\label{e.basic}
\end{equation}
where $\xbj = Q^2/2 P \cdot q$ is the Bjorken scaling variable,
with $P$ and $q$ the target nucleon and exchanged virtual photon
momenta, respectively, and for simplicity we omit explicit
flavor dependence.
The partonic differential cross section $\diff{\hat{\sigma}}$
is expressed in terms of the corresponding scaling variable
$\xbjh = Q^2/ 2\, \hat{k} \cdot q$ for the target parton with
momentum $\hat{k}$ in the subprocess (see \fref{simpleparton} below).
The function $f(\xi;Q)$ is to be interpreted as a probability
distribution of partons with fraction $\xi = \hat{k}^+/P^+$ of the
nucleon's light-cone momentum, with extra scale dependence induced
by QCD evolution.%
\footnote{We define a four-vector $v^\mu$ in terms of light-cone
	variables as $v^\mu = (v^+, v^-, \T{v}{})$, with
	$v^\pm = (v^0 \pm v^z)/\sqrt2$.}
The first term in \eref{basic} is the end result of a sequence
of canonical approximations which increase in accuracy as $Q$
increases with fixed $\xbj$~\cite{Collins:2011qcdbook}, while the
second term represents power suppressed (``${\rm p.s.}$'') errors
that are proportional to powers of $1/Q^2$ relative to the first term.
Factorization then describes the limit of large $Q$, with $\xbj$
fixed, where these error terms can safely be ignored.

Knowing the exact value of the correction term in \eref{basic} requires
a much deeper understanding of complex QCD dynamics than what is
treated by the usual factorization. However, there are certain standard
approximations (see, {\it e.g.},~Ref.~\cite[p.~95]{Ellis:1991qj})
contributing to the error in \eref{basic} that deal only with the
external kinematics of $P$ and $q$ and have nothing specifically
to do with the dynamics of the deeply inelastic collision.
These are what we will mean by ``purely kinematical'' approximations.
The most common of these is a target mass approximation in inclusive DIS:
if the target is moving in light-cone variables with large ``$+$''
momentum and zero transverse momentum, then
	$P^\mu = (P^+,M^2/2 P^+,\T{0}{}) \approx (P^+,0,\T{0}{})$.
As will be discussed in detail below, the resulting errors are
proportional to powers of $\xbj^2 \pmass^2/Q^2$, where $\pmass$
is the target nucleon mass.

By contrast, the derivation of factorization uses approximations
on internal partonic constituents, whose exact properties depend on
complex details of QCD dynamics. The resulting error terms are
suppressed by powers of $m^2/Q^2$, where $m$ here represents any
of the scales associated with intrinsic dynamical properties of
bound state partons, such as their virtualities.
Since the factorization theorem is meant to describe the limiting
behavior as $1/Q^2 \to 0$, the $\xbj^2 \pmass^2/Q^2$ errors from
the kinematical expansion are typically lumped with the dynamical
$m^2/Q^2$ errors.  We will, however, refrain from identifying the
$\order{\xbj^2 \pmass^2/Q^2}$ terms as a contribution to the
$\order{m^2/Q^2}$ corrections in all our discussions so as to
emphasize the different origins of these two types of errors.

Of course, all mass scales are ultimately fixed by the QCD scale
parameter $\Lambda_{\rm QCD}^2$, so the internal scales we associate
with $m^2$ should be understood to be proportional to $\pmass^2$:
$m^2 = \eta \pmass^2$, with $\eta$ being a dimensionless
proportionality factor.  So another way then to state the above is
that we will consider expansions in powers of $\eta \pmass^2/Q^2$
separately from powers of $\pmass^2/Q^2$.
This is explained in more detail in Secs.~\ref{s.MTA} and \ref{s.TMCs}.

At moderate $Q$, a natural question is whether all of the various
types of contributions to the error term in \eref{basic} are really
so negligible and, if not, whether some improvement is possible.
For instance, when $Q \sim 1$~GeV and $\xbj$ is not especially small
($\xbj \sim 1$), the $\xbj^2 \pmass^2/Q^2$ purely kinematical errors
may no longer be negligible.
Since they arise only from kinematical approximations, it is
reasonable to ask if these purely kinematical errors can be removed
with minimal or no modification to the basic correctness of the
factorization derivation for the first term in \eref{basic}.
In fact, as we will discuss in~\sref{TMCs}, the standard derivations
do not actually require a massless target approximation.
Setting the target mass to zero is an ancillary step, while keeping it
nonzero leads naturally to Nachtmann scaling~\cite{Nachtmann:1973mr}.
This was actually recognized some time ago by Aivazis, Olness and Tung
(AOT)~\cite{Aivazis:1993kh} in the context of heavy quark contributions
in DIS.

Questions of interpretation remain, however.  It must be established,
for example, whether it is reasonable to expect that correction for
kinematical mass errors will result in phenomenological improvements
in applications of QCD factorization.  That it should is not obvious
since there is no reason {\it a priori} to assume one type of power
correction is more important than another. The mass scales divided
by $Q^2$ that contribute errors to factorization originate from
nonperturbative features of the target hadron, so the effectiveness
of target mass improvements must be tied to specific features of
individual targets. Questions concerning the relevance of target mass
kinematics therefore cannot generally be disentangled from questions
about hadron structure.

In this paper we will argue that it is most natural to expect an
improvement from the approach of AOT~\cite{Aivazis:1993kh} if the
structure of the target involves a hierarchy of nonperturbative scales.
Keeping certain powers of $1/Q^2$ while neglecting others makes sense
only when there is a reasonably large variation in mass-squared factors
in the numerators.  Questions about the phenomenological usefulness of
kinematical target mass corrections can then be reframed as questions
about target structure.  This is how we advocate addressing the issue
of target mass kinematics more generally, as explained in more detail
in \sref{important}.
Before this, in Sec.~\ref{s.kinematics} we introduce the basic
kinematics of the DIS process at finite energy, keeping all masses
in the structure functions and the kinematic variables on which they
depend.
In Sec.~\ref{s.MTA} we introduce the massless target approximation,
carefully defining projection operators and structure functions in
the limit of small $M^2/Q^2$.
The factorization of the DIS process into a hard scattering subprocess
from massless and on-shell partons is outlined in Sec.~\ref{s.TMCs},
where we write down the explicit formulas for the structure functions
in terms of partonic scattering amplitudes and nonperturbative PDFs.
The relation between TMC improvement and nonperturbative scale
hierarchy is discussed in Sec.~\ref{s.important}.
Finally, in Sec.~\ref{s.conclusion} we summarize our results and
suggest extensions of our analysis to other applications.

\section{Deep-inelastic scattering kinematics}
\label{s.kinematics}

The reaction we will consider in the present work is inclusive
lepton scattering from a target hadron, such as a nucleon,
	\mbox{$l(l) + N(P) \to l'(l') + X(P_X)$},
where $l^\mu$ and $l'^\mu$ are the incident and scattered
lepton four-momenta, and $P^\mu$ and $P_X^\mu$ are the four-momenta
of the target nucleon and hadronic final state $X$, respectively.
The reaction will be assumed to proceed through the exchange of
a virtual photon with four-momentum $q^\mu = l^\mu - l'^\mu$.
To make the calculation more transparent, we work in a frame where
the nucleon moves in the $+z$ direction, the exchanged virtual photon
moves in the $-z$ direction, and both have zero transverse momentum.
In this case the nucleon and photon four-momenta are conveniently
parametrized in terms of light-front coordinates as
\begin{equation}
P^\mu = \parz{
        P^+,\,
	\frac{\pmass^2}{2 P^+},\,
	\T{0}{}} \, , \qquad
q^\mu = \parz{
	-\xn P^+,\,
	\frac{Q^2}{2 \xn P^+},\,
	\T{0}{}} \, ,
\label{e.qb}
\end{equation}
%
%
where $Q \equiv \sqrt{-q^2}$, and $\xn$
is the Nachtmann scaling variable~\cite{Greenberg:1972es,
Nachtmann:1973mr},
\begin{align}
\xn &\equiv -\frac{q^+}{P^+}
= \frac{2 \xbj}{1 + \sqrt{1 + 4 \xbj^2 \mhad^2/Q^2}},
\label{e.nacx}
\end{align}
so that the Bjorken variable can also be written
\begin{eqnarray}
\xbj
&=& \frac{Q^2}{2 P \cdot q}\
 =\ \frac{\xn}{(1 - \xn^2 \mhad^2/Q^2)} \, .
\label{e.xbj}
\end{eqnarray}
In the Breit frame, where the photon has zero energy, the target has
$P^+ = Q/(\sqrt{2} \xn)$ and the four-momenta simplify to
\begin{equation}
P^\mu = \parz{
        \frac{Q}{\sqrt{2} \xn},\,
	\frac{\xn \mhad^2}{\sqrt{2} Q},\,
	\T{0}{}} \, , \qquad			
q^\mu = \parz{
	-\frac{Q}{\sqrt{2}},\,
	\frac{Q}{\sqrt{2}},\,
	\T{0}{}}.
\label{e.q}
\end{equation}
The total inclusive cross section is expressed as a contraction
of leptonic and hadronic tensors,
\begin{equation}
\label{e.sidis1}
E^\prime \frac{\diff{\sigma}{} }{\diff{^3 \3{l}^\prime}{}}
= \frac{2 \alpha_{\rm em}^2 }{\parz{s - \pmass^2} Q^4} \;
  L_{\mu \nu} W^{\mu \nu},
\end{equation}
where $E'$ is the energy of the scattering lepton,
$s = (l+P)^2$ is the invariant mass squared of the system,
and $\alpha_{\rm em} = e^2/4\pi$ is the electromagnetic fine
structure constant. The leptonic tensor is 
\begin{equation}
\label{eq:lepttensor}
L_{\mu \nu}
= 2 (l_\mu l^\prime_\nu + l^\prime_\mu l_\nu
  - g_{\mu \nu} l \cdot l^\prime),
\end{equation}
and the totally inclusive hadronic tensor is defined as
\begin{align}
\label{e.hadronictensor}
W^{\mu \nu}(P,q)
&{} \equiv 4 \pi^3 \sum_X \delta^{(4)}(P + q - P_X) \,
\langle P | j^{\mu}(0) | X \rangle \langle X | j^{\nu}(0) | P \rangle .
\end{align}
Here, the $\sum_X$ symbol represents a sum over all possible final
states $| X \rangle$, including integrals
$$
\int\!\frac{\diff{^3 \3{P}_X}{}}{(2 \pi)^3 2 E_{P_X}} \, .
$$
For spin-averaged, parity-conserving scattering the hadronic tensor can
then be expanded into dimensionless structure functions according to
\begin{align}
W^{\mu \nu}
&= \parz{-g^{\mu \nu} + \frac{q^\mu q^\nu}{q^2} }
   F_1 \parz{\xbj(\xn,\pmass^2/Q^2),Q^2}			\no
&{}+ \parz{P^\mu - \frac{P \cdot q}{q^2} q^\mu}
   \parz{P^\nu - \frac{P \cdot q}{q^2} q^\nu}
   \frac{F_2\big(\xbj(\xn,\pmass^2/Q^2), Q^2\big)}{P \cdot q} \, .
\label{e.Wmunu}
\end{align}
The structure functions take all Lorentz invariants formed by $P$
and $q$ as arguments.  These include $P \cdot q$ and $Q^2$, while
independent mass dependence is left implicit.
Instead of \mbox{$P \cdot q$} we choose $\xbj$ as the independent
variable, although it turns out that $\xn$, in fact, is a more natural
choice 
in the context of factorization.
We will continue to use the Bjorken variable $\xbj$, however, since
that is the more traditional choice, but will write it in the form
$\xbj(\xn,\pmass^2/Q^2)$, as a function of $\xn$ and $\pmass^2/Q^2$
explicitly.  While this may appear cumbersome initially, it will help
make later approximation steps unambiguous.

The structure functions $F_i$ ($i=1,2$) can be calculated from
the hadronic tensor $W_{\mu\nu}$ using projection tensors,
\begin{align}
\label{e.sf_definition}
F_i\parz{\xbj(\xn,\pmass^2/Q^2),Q^2}
= {\rm P}_i^{\mu\nu}\parz{\xn,Q,\pmass}\,
  W_{\mu\nu}(P,q) \, , \qquad [i=1,2]
\end{align}
defined by
\begin{subequations}
\label{e.F12proj}
\begin{align}
\contractortot{1}^{\mu\nu} {}& \equiv
- \frac{1}{2} g^{\mu\nu}
+ \frac{2 Q^2 \xn^2}{(Q^2 + \pmass^2 \xn^2)^2} P^\mu P^\nu\ ,
\label{e.F1proj}                         \\
\contractortot{2}^{\mu\nu} {}& \equiv
  \frac{12 Q^4 \xn^3 \left(Q^2-\pmass^2 \xn^2\right)}
       {\left(Q^2 + \pmass^2 \xn^2\right)^4}
	\bigg( P^\mu P^\nu
	     - \frac{\left(Q^2 + \pmass^2 \xn^2\right)^2}{12 Q^2 \xn^2}
		g^{\mu\nu}
	\bigg) \, .
\label{e.F2proj}
\end{align}
\end{subequations}
%
%
Up to this point all of the expressions for the cross sections and
structure functions are for exact kinematics.  In the next section
we consider the limit in which the mass of the target is taken to be
much smaller than the scale $Q$, $M/Q \ll 1$.

\section{Massless Target Approximation (MTA)}
\label{s.MTA}

Purely kinematical approximations are those which can be defined in
the context of~\sref{kinematics}; that is, by considering only overall
external momentum and with no reference to hadrons' constituents or
other dynamical properties.  A kinematical approximation replaces
$P$ and $q$, and the \emph{arguments} of the structure functions
$F_i\big(\xbj(\xn,\pmass^2/Q^2),Q^2\big)$, by different, approximated
quantities, \emph{without} changing anything about the functions in
\eref{Wmunu} themselves.

Let us define the natural approximate target hadron four-momentum
$\widetilde{P}$ in a frame where it is moving at relativistic speeds
by setting the target mass to zero,
\begin{equation}
P \to \widetilde{P} \equiv (P^+,0,\T{0}{})\, .
\label{e.theapp}
\end{equation}
The massless target approximation (MTA) is the kinematical approximation
defined by the replacement
\begin{equation}
P \cdot q\ \to\ \widetilde{P} \cdot q \, , \nonumber
\end{equation}
wherever this occurs in \eref{hadronictensor}.
To set up the approximation, it is convenient to first switch
the structure function decomposition to a basis that uses
$\widetilde{P}$ instead of $P$,
\begin{align}
W^{\mu \nu}
&= \parz{-g^{\mu \nu} + \frac{q^\mu q^\nu}{q^2} }
   \widetilde{F}_1 \parz{\xbj(\xn,\pmass^2/Q^2),Q^2}	\no
&{}+ \parz{\widetilde{P}^\mu - \frac{\widetilde{P} \cdot q}{q^2} q^\mu}
   \parz{\widetilde{P}^\nu - \frac{\widetilde{P} \cdot q}{q^2} q^\nu}
   \frac{\widetilde{F}_2\big(\xbj(\xn,\pmass^2/Q^2), Q^2\big)}
	{\widetilde{P} \cdot q} \, .
\label{e.Wmunutilde}
\end{align}
Here we have defined
\begin{equation}
\widetilde{F}_i\parz{\xbj(\xn,\pmass^2/Q^2),Q^2}\
\equiv\ \widetilde{\rm P}_i^{\mu\nu}\ W_{\mu \nu}\, ,
	\hspace*{1.5cm} [i=1,2]
\label{e.contractorstotapp}
\end{equation}
with the corresponding tensors to project out the structure functions
defined by
\begin{subequations}
\label{e.tilcon}
\begin{align}
\widetilde{\rm P}_1^{\mu\nu}
&\equiv {\rm P}_1^{\mu\nu}\big(\xn,Q,0\big)
 = -\frac{1}{2}
   \bigg( g^{\mu\nu}
	- \frac{4 \xn^2}{Q^2} \widetilde{P}^\mu \widetilde{P}^\nu
   \bigg)\, ,
\label{e.tilcona}	\\
\widetilde{\rm P}_2^{\mu\nu}
&\equiv {\rm P}_2^{\mu\nu}\big(\xn,Q,0\big)
 = - \xn
   \bigg( g^{\mu\nu}
	- \frac{12 \xn^2}{Q^2} \widetilde{P}^\mu \widetilde{P}^\nu
   \bigg)\, .
\label{e.tilconb}
\end{align}
\end{subequations}
This is a more convenient basis if we ultimately want to neglect
the minus component of $P$.  Note that it is $\xn$ that appears
in the factors on the right side of Eqs.~(\ref{e.tilcon}), and
not $\xbj$.
To relate structure functions in the two bases, we use 
\begin{equation}
  \parz{\widetilde{\rm P}_i^{\mu\nu}\
  W_{\mu \nu}}_\text{\eref{Wmunu}}
= \parz{\widetilde{\rm P}_i^{\mu\nu}\
  W_{\mu \nu}}_\text{\eref{Wmunutilde}} \, .
\end{equation}
Applying the projectors (\ref{e.tilcon}) gives
\begin{subequations}
\label{e.bastrans}
\begin{align}
\widetilde{F}_1\parz{\xbj(\xn,\pmass^2/Q^2),Q^2}
       &{}= F_1\parz{\xbj(\xn,\pmass^2/Q^2),Q^2} , \, \\
\widetilde{F}_2\parz{\xbj(\xn,\pmass^2/Q^2),Q^2}
       &{}= \frac{\left(Q^2 + \pmass^2 \xn^2\right)^2}
	         {Q^2 \left(Q^2-\pmass^2 \xn^2\right)} \,
	    F_2\parz{\xbj(\xn,\pmass^2/Q^2),Q^2} \, .
\end{align}
\end{subequations}
We stress that no approximation has been made in the discussion
up to this point.
The coefficients in front of the structure functions in
Eqs.~(\ref{e.bastrans}) are, in fact, the same as those in the
literature that are referred to as ``$\xi$-scaling''
\cite{Aivazis:1993kh, Kretzer:2002fr, Schienbein:2007gr,
Accardi:2008ne, Brady:2011uy}.
The first step in the MTA is the replacement of $\xbj(\xn,\pmass^2/Q^2)$
by $\xbj(\xn,0)$ in the structure functions in \eref{Wmunutilde},
\begin{eqnarray}
W^{\mu\nu}\ \stackrel{\rm MTA}{\longrightarrow}\ \widetilde{W}^{\mu\nu}
&=&
  \bigg( - g^{\mu \nu} + \frac{q^{\mu} q^{\nu}}{q^2} \bigg)
  \widetilde{F}_1\parz{\xbj(\xn,0),Q^2} \, 	\nonumber\\
&+&
  \bigg( \widetilde{P}^\mu - \frac{\widetilde{P} \cdot q}{q^2} q^\mu \bigg)
  \bigg( \widetilde{P}^\nu - \frac{\widetilde{P} \cdot q}{q^2} q^\nu \bigg)
  \frac{\widetilde{F}_2\parz{\xbj(\xn,0),Q^2}}{\widetilde{P} \cdot q} \, ,
\label{e.incstructdecapp}
\end{eqnarray}
where $\widetilde{W}^{\mu\nu}$ is the approximate hadronic tensor.
In this approximation, \eref{xbj} gives
\begin{equation}
\xbj(\xn,0) = \xn \, ,
\end{equation}
so that $\xbj$ and $\xn$ are interchangeable in the MTA.%
\footnote{Note that an alternative way to project the
$\widetilde{F}_i$ structure functions in both
Eqs.~(\ref{e.Wmunutilde}) and (\ref{e.incstructdecapp})
is to replace the explicit $q$ vectors by
$q \to \widetilde{q} \equiv (-\xbj P^+,Q^2/(2 \xbj P^+), \T{0}{})$
and use $\xbj(\xn,0)$ in Eqs.~(\ref{e.tilcon}) instead of
$\xbj(\xn,\pmass^2/Q^2)$.  We do not do this here since
we wish to regard the $q$ vector as exact.}

The above discussion suggests a definition for the target mass
approximated structure functions $\xfunc_i$,
\begin{equation}
\xfunc_i\parz{\xbj,Q^2}
\equiv \widetilde{F}_i\parz{\xbj(\xn,0),Q^2} \, ,
\label{e.xfun}
\end{equation}
where the script notation is a shorthand that means
$\xbj(\xn,\pmass^2/Q^2)$ is understood to be everywhere
replaced by $\xbj(\xn,0)$, so that kinematical dependence
on the ratio $\pmass^2/Q^2$ is neglected.
Part of the MTA is to approximate structure functions defined in the
``tilde'' [Eq.~(\ref{e.Wmunutilde})] and
``non-tilde'' [Eq.~(\ref{e.Wmunu})] bases as being the same.
From Eqs.~(\ref{e.bastrans}), this also introduces only an
$\order{\pmass^2/Q^2}$ error.
Expanding the structure functions in powers of $M^2/Q^2$
gives a concise expression of the MTA,
\begin{align}
F_i\parz{\xbj(\xn,\pmass^2/Q^2),Q^2}
&{}= \widetilde{F}_i\parz{\xbj(\xn,\pmass^2/Q^2),Q^2}
   + \order{\frac{\xbj^2 \pmass^2}{Q^2}}  \no
&{}= \widetilde{F}_i\parz{\xbj(\xn,0),Q^2}
   + \order{\frac{\xbj^2 \pmass^2}{Q^2}}  \no
&{}= \xfunc_i\parz{\xbj,Q^2}
   + \order{\frac{\xbj^2 \pmass^2}{Q^2}} \, ,
\label{e.MTAsum}
\end{align}
where the approximation is to drop all the $\xbj^2 \pmass^2/Q^2$
errors.  In other words, assuming an exact hadronic tensor in
\eref{hadronictensor}, the MTA
[\erefs{contractorstotapp}{incstructdecapp}] is equivalent to
a set of natural argument replacements that are reasonable when $Q$ is
very large or $\xbj$ is very small.
This approximation is usually made implicitly in discussions of high
energy scattering in the literature~\cite{Ellis:1991qj}; here we have
made it very explicit so that it will be straightforward to reverse it.
Each step in \eref{MTAsum} can be traced back to the unapproximated
hadronic tensor and structure functions.  Operationally, it is
implemented by the replacement in \eref{incstructdecapp}.

This completes our general discussion of the exact and target mass
approximated structure functions, based on considerations of external
kinematics alone.  In the remainder of the paper we will specialize the
discussion to the role of the target mass in collinear factorization.

\section{The MTA and Collinear Factorization}
\label{s.TMCs}

In this section we discuss how the MTA of the last section, combined
with the standard factorization steps~\cite{Collins:2011qcdbook}, leads
to the well-known collinear factorization theorem of \eref{basic}.
Again, we will present the steps in greater detail than is common in
the literature, which will help later to unravel the source of purely
kinematical mass sensitivity.

Before any factorization approximations are made, the exact parton
momentum $k$ can in general have both a virtuality and transverse
momentum,
\begin{equation}
k = \parz{\xi P^+,\frac{k^2+\Tscsq{k}{}}{2 \xi P^+},\T{k}{}} \, .
\label{e.kexact}
\end{equation}
The steps to obtain factorization approximate certain internal lines
by exactly light-like ones.  In particular, all lines entering and
exiting the hard partonic scattering subprocess in \fref{simpleparton}
are taken to be massless and on-shell, so that in \eref{kexact} both
$|k^2|$ and $\Tscsq{k}{}$ can be taken to be $\sim \order{m^2} \ll Q^2$
and hence dropped.
The approximated parton momentum, $\hat{k}$, is then parallel to the
hadron momentum,
\begin{equation}
\hat{k} = \parz{\xi P^+,0,\T{0}{}} \, , \label{e.khat}
\end{equation}
where $\xi = \hat{k}^+/P^+$ is the fraction of the target momentum
carried by the struck parton.
These steps for approximating the partonic momenta are justified
in the standard derivations of collinear factorization, as discussed
for instance in Ref.~\cite{Collins:2011qcdbook}.
The factorization approximations make no reference to the target mass,
so none of the approximations of the previous section are necessary
to move forward with a factorization derivation.

\begin{figure}
\centering
  \begin{tabular}{c}
    \includegraphics[scale=0.6]{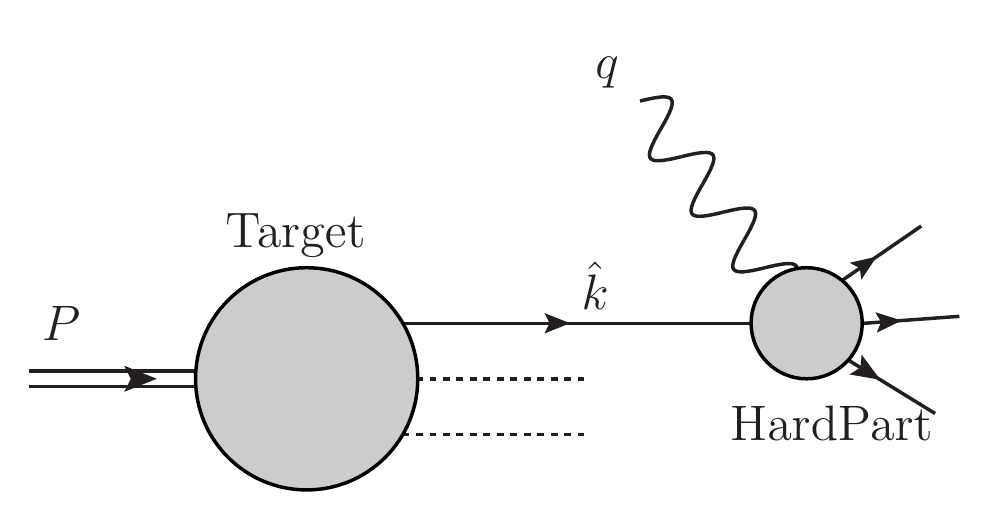}
  \end{tabular}
\caption{Illustration of DIS from a composite target ($P$) in
	collinear factorization, with hard scattering of a virtual
	photon ($q$) from an on-shell, massless parton ($\hat{k}$).}
\label{f.simpleparton}
\end{figure}

The structure tensor for the target parton in the factorized subprocess
has a form similar to that of \eref{Wmunu}, but with $P^\mu$ replaced
by $\hat{k}^\mu$,
\begin{align}
\widehat{W}^{\mu \nu}(\hat{k},q)
&{}= \parz{-g^{\mu \nu} + \frac{q^\mu q^\nu}{q^2} }
     \widehat{F}_1 \big( \xbjh(\xnh,\hat{k}^2/Q^2),Q^2 \big)	\no
&{}+ \parz{\hat{k}^\mu - \frac{\hat{k} \cdot q}{q^2} q^\mu}
     \parz{\hat{k}^\nu - \frac{\hat{k} \cdot q}{q^2} q^\nu}
     \frac{\widehat{F}_2 \big( \xbjh(\xnh,\hat{k}^2/Q^2), Q^2 \big)}
	  {\hat{k} \cdot q}\, ,
\label{e.Wmunuk}
\end{align}
where $\widehat{F}_i$ are the corresponding structure functions
for the parton.
In analogy with the scaling variables for the hadron, here
$\hat{x}_{\rm N}$ is the partonic version of the Nachtmann variable
$\xn$, as the natural generalization of \eref{nacx},
\begin{equation}
\hat{x}_{\rm N}
= - \frac{q^+}{\hat{k}^+}
= \frac{2 \xbjh}{1 + \sqrt{1 + 4 \xbjh^2 \hat{k}^2/Q^2}}
= \frac{\xn}{\xi} \, ,
\end{equation}
and $\xbjh$ is the obvious generalization of \eref{xbj},
\begin{equation}
\xbjh \equiv \frac{Q^2}{2 \hat{k} \cdot q}
= \frac{Q^2}{2 \hat{k}^+ q^-}
= \frac{\xn}{\xi} \, .
\label{e.xhatdef}
\end{equation}
Since for massless partons $\hat{k}^2 = 0$, the MTA is automatic
for the partonic structure tensor, and $\xnh = \xbjh$.
Using the notation of \eref{xfun}, but now for the partonic target,
the partonic structure tensor can be written as
\begin{align}
\widehat{W}^{\mu \nu}(\hat{k},q)
&{}= \parz{-g^{\mu \nu} + \frac{q^\mu q^\nu}{q^2} }
     \widehat{\xfunc}_1 \parz{\xbjh,Q^2}			\no
&{}+ \parz{\hat{k}^\mu - \frac{\hat{k} \cdot q}{q^2} q^\mu}
     \parz{\hat{k}^\nu - \frac{\hat{k} \cdot q}{q^2} q^\nu}
     \frac{\widehat{\xfunc}_2\parz{\xbjh, Q^2}}{\hat{k} \cdot q} \, ,
\label{e.Wmunuk2}
\end{align}
where $\widehat{\xfunc}_i$ are the partonic versions of the
massless structure functions of \eref{xfun}.
The factorization theorem, \eref{basic}, now in terms of hadronic
and partonic structure tensors, can be represented as
\begin{equation}
W^{\mu \nu}(P,q)
= \int_{\xi_{\rm min}}^1 \frac{\diff{\xi}}{\xi}\,
  \widehat{W}^{\mu \nu}\big( \hat{k}(\xi),q \big)\, f(\xi)\
+\ \order{m^2/Q^2} \, . \label{e.Wtens}
\end{equation}
For brevity here we have suppressed the dependence on the
renormalization group scale $Q$ in the PDF $f(\xi)$, but have
included the explicit $\xi$ argument of $\hat{k}(\xi)$ to
emphasize that the plus component of the target parton is
related to the hadron through the momentum fraction $\xi$.
Applying the projectors in Eqs.~(\ref{e.F12proj}) allows
factorization to be written in terms of structure functions,
still without the MTA,
\begin{subequations}
\begin{align}
F_1\parz{\xbj(\xn,\pmass^2/Q^2),Q^2}
&{}= \int_{\xi_{\rm min}}^{1} \frac{\diff{\xi}}{\xi}\,
     \widehat{\xfunc}_1\big( \xbjh(\xi),Q^2 \big)\, f(\xi)\,
   +\, \order{m^2/Q^2} \, ,  \label{e.f1p} \\
F_2\parz{\xbj(\xn,\pmass^2/Q^2),Q^2}
&{}= \frac{Q^2 \left(Q^2-\pmass^2 \xn^2\right)}
	  {\left(Q^2 + \pmass^2 \xn^2\right)^2}
     \int_{\xi_{\rm min}}^1 \diff{\xi}\,
     \widehat{\xfunc}_2\big( \xbjh(\xi),Q^2 \big)\, f(\xi)\,
   +\, \order{m^2/Q^2} \, ,  \label{e.f2p}
\end{align}
\end{subequations}
where from \eref{xhatdef} one has $\xbjh(\xi) = \xn/\xi$.
For the lower limit of the $\xi$ integration, the minimum $\xi$
occurs when $(\hat{k} + q)^2 = 0$, which gives $\xi_{\rm min} = \xn$.
Thus, without kinematical target mass approximations, the factorized
expressions for the structure functions are
\begin{subequations}
\label{e.f12pp}
\begin{align}
F_1\parz{\xbj(\xn,\pmass^2/Q^2),Q^2}
&{}= \int_{\xn}^{1} \frac{\diff{\xi}}{\xi}\,
     \widehat{\xfunc}_1(\xn/\xi,Q^2)\, f(\xi)
   + \order{m^2/Q^2}				\no
&{}\equiv F^{\rm AOT}_1 \parz{\xbj(\xn,\pmass^2/Q^2),Q^2}
   + \order{m^2/Q^2} \, ,  \label{e.f1pp} \\
F_2 \parz{\xbj(\xn,\pmass^2/Q^2),Q^2}
&{}= \frac{Q^2 \left(Q^2-\pmass^2 \xn^2\right)}
	   {\left(Q^2 + \pmass^2 \xn^2\right)^2}
     \int_{\xn}^{1} \diff{\xi}\,
     \widehat{\xfunc}_2(\xn/\xi,Q^2)\, f(\xi)
   + \order{m^2/Q^2}				\no
&{}\equiv F^{\rm AOT}_2 \parz{\xbj(\xn,\pmass^2/Q^2),Q^2}
   + \order{m^2/Q^2} \, .  \label{e.f2pp}
\end{align}
\end{subequations}
The errors here arise entirely from assumptions about the smallness
of intrinsic parton scales; there are no $\xbj^2 \pmass^2/Q^2$ types
of errors since no MTA has been made.
The second lines of Eqs.~(\ref{e.f1pp}) and (\ref{e.f2pp})
define the ``AOT structure functions'', $F^{\rm AOT}_i$, as the
factorized structure functions with exact external kinematics
\cite{Aivazis:1993kh}, and this prescription for taking target
masses into account will be referred to as the AOT method.
(Note that the notation in Eqs.~(\ref{e.f12pp}) differs from that
in Ref.~\cite{Aivazis:1993kh}, whose focus was more on the treatment
of heavy quark effects rather than on kinematical errors.)
If, in addition, $\xn$ is expanded in powers of
$\xbj^2 \pmass^2/Q^2$, then Eqs.~(\ref{e.f12pp}) become
\begin{subequations}
\label{e.f12ex}
\begin{align}
F_1 \parz{\xbj(\xn,\pmass^2/Q^2),Q^2}
&{}= \xfunc_1(\xbj,Q^2)
   + \order{\frac{\xbj^2 \pmass^2}{Q^2}}		\no
&{}= \int_{\xbj}^{1} \frac{\diff{\xi}}{\xi}\,
     \widehat{\xfunc}_1(\xbj/\xi,Q^2)\, f(\xi)
   + \order{{\rm max}
     \left[ \frac{m^2}{Q^2} \, ,\frac{\xbj^2 \pmass^2}{Q^2}
     \right] } \, ,  \label{e.f1ex} \\
F_2 \parz{\xbj(\xn,\pmass^2/Q^2),Q^2}
&{}= \xfunc_2(\xbj,Q^2)
   + \order{\frac{\xbj^2 \pmass^2}{Q^2}}		\no
&{}= \int_{\xbj}^{1} \diff{\xi}\,
     \widehat{\xfunc}_2(\xbj/\xi,Q^2)\, f(\xi)
   + \order{{\rm max}
     \left[ \frac{m^2}{Q^2} \, ,\frac{\xbj^2 \pmass^2}{Q^2}
     \right] } \, .  \label{e.f2ex}
\end{align}
\end{subequations}

The expressions in Eqs.~(\ref{e.f12pp}) are the most immediate
results of a factorization derivation of the style of
Ref.~\cite{Collins:2011qcdbook}, and the factorized terms on the
right-hand-side can be considered nearly exact if the $m^2/Q^2$ errors
({\it i.e.}, quantities like parton virtuality) are negligible.
On the other hand, Eqs.~(\ref{e.f12ex}) are the more usual way of
presenting the final factorization result, which arises from
applying the MTA of Sec.~\ref{s.MTA} to the factorized expressions
in Eqs.~(\ref{e.f12pp}).  The resulting errors are suppressed by
$\xbj^2 \pmass^2/Q^2$ and are here seen to be of purely kinematical
origin.
The approximation of dropping all power corrections in \eref{f12ex}
and keeping only the first term on the right will be referred to as
the ``factorized massless target approximation'' (FMTA), since it
just combines standard factorization with the MTA.  If we wish to
keep kinematical target mass effects, we will simply maintain
Eqs.~(\ref{e.f12pp}).

In order to make the various approximations very
explicit, the discussion in the last two sections of the basic
theoretical set up has been much more detailed than what is usually
found in the literature.  This has required the introduction of a
number of new notations for structure functions, which is useful
to briefly summarize here:
\begin{itemize}
\item	Hadronic structure functions, which are represented by
	the Roman font $F_i$, are functions of the
	independent variables $\xbj$ and $Q^2$; however, since
	it is ultimately convenient to express them in terms of
	$\xn$ and $Q^2$, we write $\xbj$ explicitly as a function
	of $\xn$ and $M^2/Q^2$ as in \eref{Wmunu}.
\item	The hadronic tensor can be re-expressed in a different
	basis of Lorentz vectors, by using $\widetilde{P}^\mu$
	rather than $P^\mu$ to define the corresponding structure
	functions $\widetilde{F}_i$ in the massless basis, which
	we distinguish by the tilde [``\ \ $\widetilde{}$\ \ ''] symbol.
\item	When this is combined with the approximation
	$\xbj(\xn,\pmass^2/Q^2) \to \xbj(\xn,0)$ we obtain the
	$\widetilde{F}_i\parz{\xbj(\xn,0),Q^2}$ structure
	functions evaluated as in \eref{incstructdecapp}.
\item	The script notation for the structure functions	$\xfunc_i$
	is an abbreviation for the special case when $\pmass^2/Q^2$
	is set to zero in $\xbj(\xn,\pmass^2/Q^2)$, as in \eref{xfun}.
\item	A hat [``\ \ $\widehat{}$\ \ ''] on a structure function
	denotes a massless and on-shell partonic target.
	Note that structure functions in Roman font with a hat
	($\widehat{F}_i$) and in script font with a hat
	($\widehat{{\cal F}}_i$) are identical, since
	$\hat{k}^2 = 0$.
	Also, partonic structure functions are identical with
	(the partonic analogues of) either the
	$W^{\mu\nu}$ [\eref{Wmunu}] or
	$\widetilde{W}^{\mu\nu}$ [\eref{incstructdecapp}]
	bases, since the target parton in the hard part is
	always massless and on-shell.
\end{itemize}
For many subsequent practical applications some of these notations
will be redundant; however, since they make the different layers of
conventions and approximations very explicit, they will be useful
for our present purposes.

To conclude this section, let us also summarize the key observations:
\begin{enumerate}
\item[(1)] There are two independent types of approximations.
	One is the purely kinematical approximation described
	in \sref{MTA}, with errors suppressed by powers of
	$\xbj^2 \pmass^2/Q^2$.  It is independent of whatever
	theoretical techniques might be used to actually
	calculate the structure functions.
	The second approximation is the factorization theorem
	in \eref{Wtens}, with errors suppressed by powers of
	$m^2/Q^2$, where $m^2$ is a typical scale associated
	with intrinsic dynamical properties of partons,
	such as their virtualites.
\item[(2)] The MTA is not necessary for deriving collinear
	factorization.  The relation $\xbjh = \xn/\xi$ in
	\eref{xhatdef} is usually automatically	approximated
	to $\xbj/\xi$, but this is not needed.  One may simply
	stop at Eqs.~(\ref{e.f12pp}) and view the MTA application
	that leads to Eqs.~(\ref{e.f12ex}) as ancillary.
\item[(3)] The standard factorization derivation, as embodied in the
	AOT method, automatically gives $\xn$ instead of $\xbj$ as
	the natural scaling variable for the structure functions
	(neglecting logarithmic $Q$ dependence from higher orders
	in $\alpha_s$).
\end{enumerate}

Before concluding, let us also mention that a number of other
prescriptions for dealing with the effects of a nonzero target mass
on kinematics have been proposed in the literature, but generally
these impose extra assumptions on the dynamics.  We discuss these
in more detail in \aref{appendix}.
Having reviewed the mathematical statement of factorization in the
presence of target masses in detail, and the corresponding expressions
for the structure functions, in the next section we turn to the
question of the physical interpretation of an observed improvement
from target mass effects.

\section{When are Target Mass Kinematics Relevant?}
\label{s.important}

The most straightforward and correct approach to computing the
inclusive DIS structure functions is to simply avoid introducing
unnecessary kinematical errors by choosing to keep target
momentum exact and applying the AOT expressions for factorization
in Eqs.~(\ref{e.f12pp}).
A~question of interpretation remains, however; without special
knowledge of the target structure there is no reason {\it a priori}
to expect the powers of $\xbj^2 \pmass^2/Q^2$ from purely
kinematical approximations to be any more important than other
power-suppressed corrections.

\subsection{Scattering from subsystems}
\label{ss.subsystems}

To interpret an observed phenomenological improvement obtained
by using the AOT method instead of the FMTA, consider several
generic scenarios for scattering from an extended target that
could reveal a nontrivial relation between target mass effects
and general properties of hadron structure.
Consider, for instance, that if the target is a composite object
(the precise nature of which need not be specified at this stage),
then the sum of scattering amplitudes may described as occurring
off subsystems of the target, as depicted in \fref{pictures}. 
We leave the nature of the dynamics completely unspecified at this
stage and only assume that diagrammatic arguments apply generally.
To be completely general, we also allow for the possibility that
the lower (nonperturbative) blob is empty so that scattering can
occur off the entire target as a whole.

To be quantitative, we define the generic subsystem to have a
momentum before the collision parametrized by the four-vector
\begin{equation}
p = \parz{X P^+, \frac{\mtsq}{2 X P^+}, \T{p}{}} \, ,
\label{e.pparam}
\end{equation}
where the squared transverse mass $\mtsq \equiv p^2 + \Tscsq{p}{}$
denotes the sum of the virtuality $p^2$ (which could in principle be
negative) and transverse momentum $\Tscsq{p}{}$ of the subsystem,
and $X = p^+ / P^+$ is the light-cone fraction of the target
carried by the subsystem.
The collision with the exchanged virtual photon produces another
system of particles with invariant mass-squared
\begin{equation}
\finmass^2
\equiv (p + q)^2 \, .
\end{equation}
Such a system need not be physical and could be off-shell;
for example, it could be a part of a hadronizing string.
Without loss of generality, we may describe the total lepton scattering
amplitude for the whole target $\mathcal{A}^{\rm tot}(P,q,l')$, which
in general depends on three variables (chosen here to be $P$, $q$ and
$l'$), in terms of the amplitude for scattering off the subsystem,
	$$ \mathcal{A}^{p}(p,q,l'). $$
To connect to the total amplitude $\mathcal{A}^{\rm tot}$, the
subsystem amplitude needs to be integrated over all components of
$p$, weighted by a function that characterizes the four-momentum
distribution of the subsystem in the overall target.

\begin{figure}[t]
\centering
\includegraphics[scale=0.5]{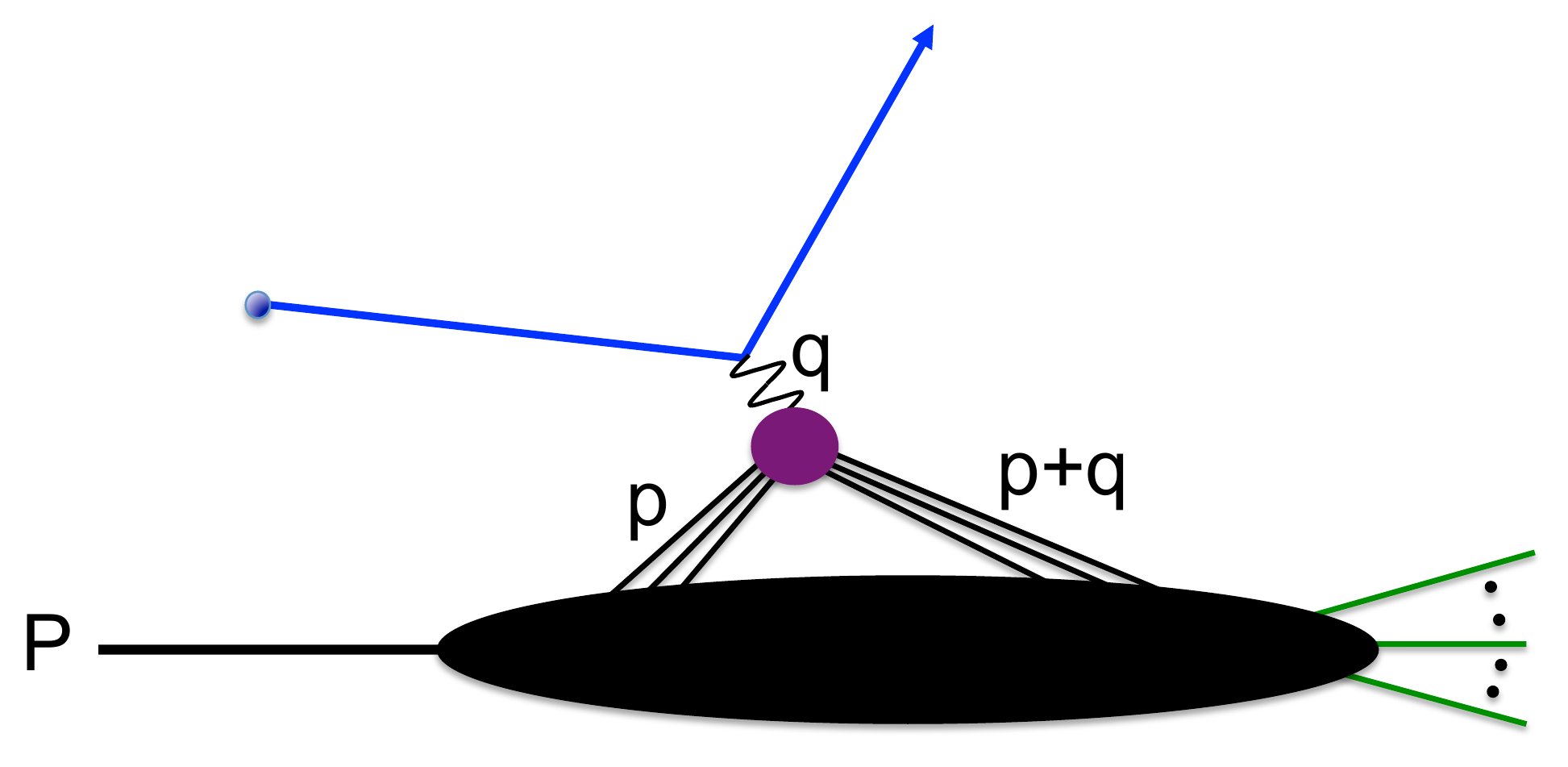}
\caption{DIS from a subsystem ($p$) of a composite target ($P$).
	The solid lines connecting to the virtual photon ($q$)
	through the upper blob can be any constituents of the target.}
\label{f.pictures}
\end{figure}

To avoid confusion in what follows below, it is important not to view
the diagram in \fref{pictures} as the sort of ``region'' diagram common
in factorization derivations~\cite{Collins:2011qcdbook}, but rather as
a topological representation in which the blobs are not necessarily
characterized by any particular (small or large) momentum.
The blobs simply denote an arbitrary subgraph assignment for some
graphical contribution to the amplitude; some lines are routed through
the (upper) photon--subsystem part of the graph, while others are
diverted through the (lower) part of the graph connected to the
target.

Such organization does not achieve much of interest until we pose
questions about possible relationships between the total target
and subsystem momenta, $P$ and $p$.  If we find that there is an
assignment in \fref{pictures} such that
	$\Tscsq{p}{}, \mtsq \ll Q^2$
for typical values of $\Tscsq{p}{}$ and $\mtsq$, then up to
power-suppressed errors the amplitude for scattering from the
subsystem becomes a function of $X$ only,
\begin{equation}
  \mathcal{A}^{p}(p,q,l')
= \mathcal{A}^{p}(X,q,l')
+ \order{\frac{m^2}{Q^2}} \, ,  \label{e.xsecexp}
\end{equation}
where $m^2$ refers to $\Tscsq{p}{}$ or $\mtsq$.
The entire factorization derivation can then be performed for the
sub-amplitude $\mathcal{A}^{p}(X,q,l')$ rather than for
the total amplitude $\mathcal{A}^{\rm tot}(P,q,l')$.

In general the invariant mass $\finmass^2$ varies between small
values ($\approx 0$) and large values (of order $Q^2$ or larger).
In the standard QCD factorization paradigm, large-$\finmass^2$
behavior is describable by perturbative calculations.  One can
therefore define an approximate invariant mass squared $\finmassex^2$
of the final state subsystem which is calculated by approximate
methods that deal with values of $\finmass^2/Q^2 = \order{1}$,
\begin{equation}
\finmass^2 \equiv \finmassex^2 + \findelt^2,
\end{equation}
where $\findelt^2$ is the correction needed to recover the exact
$\finmass^2$ value.
The approximate invariant mass squared $\finmassex^2$ may vary
from zero to $\order{Q^2}$, while $\findelt^2$ is of the order
of a typical small scale comparable to $\Tscsq{p}{}$ and $\mtsq$.
Expanding $X$ in terms of these variables, we can write
\begin{subequations}
\label{e.expansion}
\begin{align}
\xsub
&= \xn
   \left\{1 + \frac{\finmassex^2}{Q^2}
	+ \frac{\Tscsq{p}{} + \findelt^2}{Q^2}
	- \frac{\mtsq (\Tscsq{p}{} + \finmassex^2 + \findelt^2)}{Q^4}
	+ \cdots
   \right\},
\label{e.expansion_xn}
\end{align}
and, further expanding the Nachtmann variable $\xn$, the light-cone
fraction becomes
\begin{align}
\xsub
&= \xbj
   \left\{1 + \frac{\finmassex^2}{Q^2}
	+ \frac{\Tscsq{p}{} + \findelt^2 - \xbj^2 \pmass^2}{Q^2}
	- \frac{(\mtsq + \xbj^2 \pmass^2)
	        (\Tscsq{p}{} + \finmassex^2 + \findelt^2)
		- 2 \xbj^4 \pmass^4}
	       {Q^4}
	+ \cdots
   \right\} \, .
\label{e.expansion_xb}
\end{align}
\end{subequations}
If the typical values of small mass scales associated with the
interactions between subsystems ($\Tscsq{p}{}$, $\mtsq$ and
$\findelt^2$) are totally negligible, but $\xbj^2 \pmass^2$ is
comparatively large, then the expansion in \eref{expansion_xn}
is an improvement over the expansion in \eref{expansion_xb}.
In other words, in the limit of large $Q$,
\begin{subequations}
\label{e.approxX}
\begin{equation}
X\ \approx\ \xn \parz{ 1 + \frac{\finmassex^2}{Q^2} } \,
	\label{e.improveX}
\end{equation}
provides a better approximation than
\begin{equation}
\ \
X\ \approx\ \xbj \parz{ 1 + \frac{\finmassex^2}{Q^2}  } \, .
	\label{e.badX}
\end{equation}
\end{subequations}
In both of these cases, the connection between $X$ and external
observables has lost any sensitivity to the details of interactions
between subsystems.  The only dependence on dynamics is through
$\finmassex^2$, which is calculable in factorization and perturbation
theory.  Suggestively defining
\begin{equation}
\parz{ 1 + \frac{\finmassex^2}{Q^2} }\, \equiv\, \frac{1}{\xi} \, ,
\end{equation}
the subsystem amplitude in \eref{xsecexp} can be written
\begin{subequations}
\label{e.xsecexp1}
\begin{align}
\mathcal{A}^{p}(p,q,l')
&{}= \mathcal{A}^{p}(\xn/\xi,q,l')
   + \order{\frac{m^2}{Q^2}} \,			\label{e.xsecexp2} \\
&{}= \mathcal{A}^{p}(\xbj/\xi,q,l')
   + \order{{\rm max} \left[ \frac{m^2}{Q^2} \, ,\frac{\xbj^2 \pmass^2}{Q^2}
		      \right] } \, .		\label{e.xsecexp3}
\end{align}
\end{subequations}
If
	$\xbj^2 \pmass^2 \sim Q^2$
but
	$\Tscsq{p}{}, \mtsq, \findelt^2 \ll Q^2$,
then truncating the expansion in (\ref{e.xsecexp2}) is valid
while in (\ref{e.xsecexp3}) it is not.
If, however,
	$\xbj^2 \pmass^2 \sim \Tscsq{p}{}, \mtsq, \findelt^2$,
then there is no reason to expect either expansion to be any better
or worse than the other.  The same statements apply to the overall
cross section, since it is related to $\mathcal{A}^{p}$ by
taking the square modulus, summing over hadronic final states,
and integrating over $\T{p}{}$ and $m_{\rm T}$ (whose typical values
are restricted by the $\Tscsq{p}{}, \mtsq \ll Q^2$ assumption to be
small and are thus decoupled from $\mathcal{A}^{p}$).

The above discussion naturally leads us to the conclusion that,
if $\xbj^2 \pmass^2$ is large but subsystem scales are small,
then the cross section reduces to a function of $\xn/\xi$,
with the momentum fraction $\xi$ calculable from methods that
account for large $\finmassex^2$ --- all of which can be performed
within standard factorization.  The AOT set of expressions
[Eqs.~(\ref{e.f12pp})] is just a specific realization of this
within collinear factorization.
Namely, the hard scattering subprocess is always a function of $\xn$,
while large final state invariant masses in the hard part of the
scattering amplitude are accounted for by using $\xn/\xi$ in the
subprocess, with $\xi$ obtained as in Eqs.~(\ref{e.approxX}).
In other words, if the typical $| p^2 |$ is small and $p$ is collinear
to $P$, then the steps for deriving factorization can be applied
directly to $|\mathcal{A}^{p}(p,q,l')|^2$ with $p^2 = 0$
rather than to $|\mathcal{A}^{\rm tot}(P,q,l')|^2$.
The result is automatically the AOT factorization in
Eqs.~(\ref{e.f12pp}).  Furthermore, since it accounts for large
$\finmassex^2$, the AOT improvement applies to all orders in
perturbation theory.

\subsection{TMC improvement and hierarchy of scales}
\label{ss.hierarchy}

Now we may ask what general characteristics of a composite target can
give rise to a scenario where $\Tscsq{p}{},\mtsq \ll \xbj^2 \pmass^2$,
which would justify the result in \eref{xsecexp2} being an improvement
over that in \eref{xsecexp3}.
At one extreme, it cannot be the case of scattering from a single,
isolated perturbative quark or gluon, as these can emit large amounts
of collinear and soft radiation.  Moreover, a perturbative quark has
virtuality that ranges up to $\order{Q^2}$.  
%
%
A system of collinearly propagating quarks and gluons that are 
nearly massless and on-shell cannot be described purely in terms of 
short-distance, perturbative propagators.
At the other extreme, the $\Tscsq{p}{},\mtsq \ll \xbj^2 \pmass^2$
condition also cannot arise when all or most of the lines in
\fref{pictures} are routed through the upper part of the diagram,
leaving the blob in the lower part of the diagram completely empty,
which would correspond to $m_{\rm T} \sim \pmass$.

The only way, therefore, to consistently arrive at a scenario
whereby $\Tscsq{p}{}, \mtsq \ll \xbj^2 \pmass^2$, and thus
\eref{xsecexp2} (in terms of $\xn$) be an improvement over
\eref{xsecexp3} (in terms of $\xbj$), is if the target consists of
more than one separate, low-invariant mass
(relative to $\xbj^2 \pmass^2$) subsystem that can play
the role of the lines entering the upper blob in \fref{pictures}.
To avoid pushing $| p^2 |$ too high, the interactions between 
subsystems need to be reasonably weak.
While the individual subsystems necessarily need to have a small
typical invariant mass $|p^2|$ relative to $\xbj^2 \pmass^2$,
each subsystem can involve internal interactions that involve
scales much larger than $\Tscsq{p}{}, \mtsq, \findelt^2$,
but still much smaller than $Q^2$.
Therefore, it is only the scales involved in the interactions
\emph{between} subsystems that need to be very small in order for
the above argument for the usefulness of the AOT method to be valid.

Our general conclusion is that any observed improvement in the
theoretical description of scattering that comes from using
\eref{xsecexp2} instead of \eref{xsecexp3} is suggestive of a
hierarchy of ``clustered'' structures within the target,
representing correlated subsystems of strongly interacting particles.
We stress that we are totally agnostic about what
those clusters might be; our observation is simply that,
kinematically, some sort of clustering is preferred.
Thus, an improvement in the phenomenological description using the
AOT method can be interpreted as evidence that scattering occurs off
a collection of weakly interacting subsystems (since $\Tscsq{p}{}$,
$\mtsq$ and $\findelt^2$ must be small relative to $\xbj^2 \pmass^2$),
while a failure to observe any improvement suggests a more complicated
type of scattering.
(Some of this also echoes earlier discussions of TMCs in DIS at low
energies, such as in Ref.~\cite{DeRujula:1976baf}, see pg. 325, where
the scale $M_0$ there is analogous to the mass $m$ used in the present
work.)
A subsystem can in general be any nonperturbative system,
consisting of one or more interacting particles, whose internal
interactions are stronger than interactions with other subsystems
in the target.  The subsystem could, for example, be colored or
colorless; for the latter, we notice that for a nucleon target
the region of kinematics where the $\xbj^2 M^2/Q^2$ corrections
are important corresponds to the nucleon resonance region,
and the subsystems might be a collection of hadrons, such as
nucleons and pions.
However, the exact nature of the target or its subsystems and
their interactions is not relevant to our discussion.

The above argument is very general, since it only relies on the
kinematics of scattering off subsystems in a target, and the
assumption that scattering from the composite object can be described
in generally diagrammatic terms.  In particular, it applies to
arbitrary orders in perturbation theory.
In fact, arriving at Eqs.~(\ref{e.xsecexp1}) does not even require
factorization or partonic degrees of freedom specifically.
It only states that, if scattering occurs off weakly interacting
light and nearly on-shell subsystems in a heavier target, then the
cross section at a particular $\finmass^2$ becomes a function
of $\xn/\xi$, where $\xi$ is either 1 or is obtainable from
\mbox{large-$\finmassex^2$ methods}.

An example of such a scale hierarchy could be nuclear targets,
where the subsystems correspond to nucleons; the hierarchy arises
because interactions between nucleons are much weaker than the
typical interactions binding quarks and gluons inside the
nucleons~\cite{Geesaman:1995yd, Malace:2014uea}.
Other examples may be nucleons coupled to soft pseudoscalar mesons
through chiral dynamics, which can give rise to unique nonperturbative
features in sea quarks in the proton~\cite{Thomas:1983fh,
Signal:1987gz, Salamu:2014pka, Wang:2016ndh, Geesaman:2018ixo}.
A possible hierarchy with explicit color degrees of freedom could
involve partons clustered into constituent quark-like
subsystems~\cite{Close:1979bt, Richard:2012xw}.
Conversely, an example of a target where one would \emph{not} expect an
improvement would be the case of a hadron target whose mass comes almost
entirely from a single point-like quark, such as a heavy quark hadron.
We stress again, however, that our arguments here do not rely on any
assumptions about dynamics of the composite object or the nature
of its subsystems, but only on the kinematical considerations
associated with target mass improvement.

\section{Conclusion}
\label{s.conclusion}

In this paper we have presented a detailed description of the basic
structure function analysis of deeply inelastic scattering in the
context of QCD factorization, fully taking into account hadronic masses
in order to give clarity to the notion of ``purely kinematical''
mass effects.  Even when clearly stated, however, the meaning of an
improvement in the theoretical description of the scattering process
from purely kinematical effects of the target mass begs for a physical
interpretation.

The discussions in Secs.~\ref{s.MTA}--\ref{s.important} make clear
that an improvement is natural if factorization is understood to apply 
to scattering off a small invariant mass subsystem or cluster inside a
composite target.  
Models of the nucleon with multiple scales and a clustering structure
imply a particular kind of phenomenological prediction --- that
standard collinear QCD factorization, in the form of AOT framework
for treating target masses with exact external kinematics, can be
extended to smaller $Q$ and larger $\xbj$ than might otherwise be
expected from perturbative QCD arguments.
In the limit of large $Q$, with all other scales fixed, and assuming
$\xbj \pmass \approx Q$, it is the first terms on the right hand sides
of Eqs.~(\ref{e.f1pp}) and (\ref{e.f2pp}) that give the asymptotic
behavior. 
The clustering hypothesis suggests that, as $Q$ decreases, the power
corrections initially come mainly from switching between $\xbj$ and
$\xn$ in the usual factorized expressions, and also accounting for
overall kinematic factors such as in \eref{f2pp}.

An interesting consequence is that the degree of purely kinematical 
improvement found by keeping the target mass can be viewed as probing 
the degree of clustering in the target.  To quantify this, it will be 
interesting to investigate how much improvement can be expected within
specific models of the target.  This way of viewing the target mass
effects suggests a variety of future directions for research.

From phenomenological and global QCD analyses of deep inelastic
lepton--nucleon scattering data, it is already well established that
treatments of the target mass that switch $\xbj$ to $\xn$ significantly
improve the description of the data and extend its range to lower $Q$
and larger $\xbj$ values~\cite{DeRujula:1976baf, Ji:1994br,
Bianchi:2003hi, Accardi:2009br, Owens:2012bv, Kulagin:2000yw,
Alekhin:2017fpf, Accardi:2016qay}.
On the other hand, clear room for refinement exists, for example to
distinguish between precise implementations of TMCs that have been
proposed in the literature~\cite{Georgi:1976ve, Barbieri:1976rd,
Ellis:1982cd, Isgur:2001bt, Steffens:2006ds, Schienbein:2007gr,
Accardi:2008ne, Brady:2011uy, Steffens:2012jx}.
Also, upcoming experiments will allow for comparison between
different target structures, including pions, kaons, and nuclei
\cite{Arrington:2001ni, Arrington:2003nt, Melnitchouk:2003dz,
Melnitchouk:2005zr}.
While the discussion in the present work has for simplicity
been restricted to a single flavor, the generalization to the
more realistic case of multiple flavors is straightforward.
Moreover, the treatment of structure functions in
Secs.~\ref{s.kinematics} through \ref{s.TMCs} can be directly
extended to spin and polarization dependent structure functions.
This will be important since the extraction of certain spin dependent
effects can be especially sensitive to target mass effects
\cite{Wandzura:1977ce, Matsuda:1979ad, Piccione:1997zh,
Blumlein:1998nv, Bosted:2006gp, Solvignon:2008hk}.
We leave these interesting and important topics for future
consideration.

\newpage
\begin{acknowledgments}

We thank J.~C.~Collins for useful discussions.
This work was supported by the U.S. Department of Energy, Office of 
Science, Office of Nuclear Physics, under Award Number DE-SC0018106,
and by the DOE Contract No.~DE-AC05-06OR23177, under which Jefferson
Science Associates, LLC operates Jefferson Lab.
F.S. is funded by the Deutsche Forschungsgemeinschaft (DFG) project
number 392578569.

\end{acknowledgments}

\appendix
\section{Contrast with other TMC methods}
\label{a.appendix}

Throughout this paper we have adopted what could be viewed as the
most natural meaning of a ``purely kinematical correction'';
namely, a correction that is totally independent of any assumptions
pertaining to the dynamics within the target.
The MTA from \sref{MTA} accounts for all such approximations that one
encounters in the context of standard collinear factorization in DIS.
The purely kinematical target mass correction is therefore uniquely
of the form derived by AOT~\cite{Aivazis:1993kh} (see \sref{TMCs}),
since this is merely the combination of the MTA and standard
factorization, which is independent of target mass kinematics.
Any other corrections must involve at least some set of additional
assumptions about parton dynamics.

In the literature there exist a number of other prescriptions that are
sometimes described as ``purely kinematical'' target mass corrections,
but which in various ways differ from the AOT approach.
Probably the best known of these is the treatment by Georgi and
Politzer~\cite{Georgi:1976ve} based on the operator product expansion
(OPE).
(For extensions to the polarized case see Refs.~\cite{Wandzura:1977ce,
Matsuda:1979ad, Piccione:1997zh, Blumlein:1998nv}.)
Here the expressions for target mass corrected structure functions
contain extra terms involving integrals of structure functions, which
arise from additional constraints or assumptions that are beyond
the purely kinematical corrections implicit in the AOT approach.
As discussed by Ellis, Furmanski and Petronzio~\cite{Ellis:1982cd}, and
more recently by D'Alesio, Leader and Murgia~\cite{DAlesio:2009cps},
the origin of the additional integral factors is the constraint that
the struck partons inside the target correlation function should be
\emph{exactly} massless and on-shell, for all longitudinal momenta
and for all transverse momenta.
Absent some exotic dynamical mechanisms within the target, this appears
to be a relatively strong assumption, which in itself is not a necessary
one for the standard derivation of collinear factorization.

Another way to understand the difference between the AOT approach and
the OPE-based prescription is to note that in the latter the kinematical
TMCs that are kept are only those that are relevant for a leading twist
treatment, while kinematical corrections associated with higher twists
are neglected.  This type of assessment of $\order{m^2/Q^2}$-type errors
runs the risk of entangling the $\order{\xbj^2 \pmass^2/Q^2}$ target
mass corrections with those from other sources.
By refraining from introducing $\order{\xbj^2 \pmass^2/Q^2}$-type errors
from the outset, the direct method used by AOT has the advantage of
including all kinematical target mass effects regardless of twist.
It is worth emphasizing here that modern derivations of factorization
{\it do not} need to use the OPE, but rather can be formulated as
direct, arbitrary-order expansions in powers of
$1/Q^2$~\cite{Collins:2011qcdbook}.  An added benefit of the direct
method, which can be argued to be the more rigorous one, is that it
does not {\it a priori} need to entail an MTA.

Still other TMC formalisms have been proposed that also differ from,
or go beyond, AOT~\cite{Ellis:1982cd, Accardi:2008ne}.  For example,
the Accardi-Qiu prescription~\cite{Accardi:2008ne} uses collinear
factorization together with the dynamical assumptions that
well-defined target and jet directions exist at rather low
$Q^2$~\cite{Accardi:2018gmh, Sterman:1986aj} and that the
initial state baryon number flows only along one such
direction~\cite{Guerrero:2017yvf}.  This relies on a very literal
matching between virtual partonic states and a particular final
state distribution of hadrons, which goes beyond the standard
factorization paradigm~\cite{Collins:1988gx, Collins:2011qcdbook}
but regulates the behavior near the kinematical threshold at
$\xbj = 1$.

The direct factorization approach can also help to contextualize the
so-called ``threshold problem''~\cite{Georgi:1976ve}, which is the
observation that the structure function for nonzero target mass in
the OPE derivation has support at $\xbj = 1$ (where kinematically
only elastic scattering should contribute) and can be nonzero in the
unphysical region $\xbj > 1$ (up to $\xn = 1$)~\cite{DeRujula:1976ih}.
This has led to various proposals for modifying the target mass
corrected structure functions such that they have support only
in the physical region~\cite{DeRujula:1976ih, Johnson:1979ty, 
Bitar:1978cj, Steffens:2006ds, Steffens:2012jx, DAlesio:2009cps}.
The solution to the ``threshold problem'' from the factorization
perspective is simply that the conditions for which QCD factorization
itself is valid break down as $\xbj \to 1$.  While the structure
functions are defined through Eq.~(\ref{e.sf_definition}) for all
$\xbj \leq 1$, and the parton distribution $f(\xi)$ exists for all
parton momentum fractions $\xi \in [0,1]$, the factorization formulas
in Eqs.~(\ref{e.Wtens}) and (\ref{e.f12pp}) relating the two receive
increasingly large corrections at large $\xbj$ that render the
perturbative expansion in powers of both $\alpha_s$ and $1/Q^2$
no longer a useful one.
Improvements beyond this require more sophisticated methods for
treating the large-$\xbj$ region than what is available in the
standard factorization treatment.

\newpage

\bibliography{bibliography}

\end{document}